\documentclass[aps,amsmath,amsfonts,11pt]{revtex4}
\usepackage{graphicx}
\usepackage{epstopdf}
\usepackage{epsfig,graphicx}
\usepackage[english]{babel}
\usepackage{amsfonts}
\usepackage{amsmath}
\usepackage{latexsym}
\usepackage{graphics,bm}
\usepackage{dcolumn}
\usepackage{natbib}
\usepackage{bm}
\usepackage{rotating}

\begin{document}

\title{Controllable nonlinear effects in a hybrid optomechanical  semiconductor microcavity containing a quantum dot and Kerr medium}

\author{ Sonam Mahajan$^{1}$ and Aranya B Bhattacherjee$^{2}$ }

\address{$^{1}$Department of Physics, University of Petroleum and Energy Studies, Dehradun-248007, Uttarakhand, India} \address{$^{2}$Department of Physics, Birla Institute of Technology and Science, Pilani, Hyderabad Campus,  Telangana State - 500078, India}

\begin{abstract}
We theoretically investigate the nonlinear effects in a hybrid quantum optomechanical system consisting of two optically coupled semiconductor microcavities containing a quantum dot and a Kerr nonlinear substrate.The steady state behavior of the mean intracavity optical field demonstrates that the system can be used as an all optical switch. We further investigate the spectrum of small fluctuations in the mechanical displacement of the movable distributed Bragg reflectors (DBR) and observe that normal mode splitting (NMS) takes place for high Kerr nonlinearity and pump power. In addition, we have shown that steady state of the system exhibits two possible bipartite entanglements by proper tuning of the system parameters. The entanglement results suggest that the proposed system has the potential to be used in quantum communication platform. Our work demonstrates that the Kerr-nonlinearity can effectively control the optical properties of the hybrid system, which can be used to design efficient optical devices.
\end{abstract}

\pacs{03.75.Kk,03.75.Lm, 42.50.Lc, 03.65.Ta, 05.40.Jc, 04.80.Nn}

\maketitle

\section{Introduction}

With the advancement of technology in recent years, a rapidly developing field of research has emerged in the area of quantum optomechanics, which deals with the coherent interaction of optical cavity field with a mechanical mode via radiation pressure \citep{1,2,3,4,5,YW,Zhang}. This interaction has led to many applications such as gravitational wave detection interferometers \citep{6,7}, atomic force microscopes \citep{8,9}, ultrahigh-precision measurements \citep{10}, quantum information processing \citep{11,12}, quantum entanglement \citep{13,14,15} and optomechanically induced transparency (OMIT) \citep{16,17,18,19,20,21,22,23,24}. Theoretically, an optomechanical system had been proposed to study optomechanically induced amplification and perfect transparency in a double-cavity optomechanical system indicating significant progress towards signal amplification, light storage, fast light, and slow light in quantum information processing \citep{25}.

One of the obstacles of Moor's law imposed by device miniaturization is the technological development of classical computing reaching the fundamental limit \citep{26}. Quantum Dots (QDs) seem to be perfect candidates to overcome this limitation due to their narrow linewidths and their capability of implementing optoelectronic devices with optical tunability. In addition, their coupling with microcavities provide an ideal system to study the field of cavity quantum electrodynamics. Recent studies have observed the strong coupling regime of cavity quantum electrodynamics in which the photonic mode of cavity and exciton modes of QD mix together to produce quasi-particles called polaritons \citep{27,28,29}. These kind of strong interactions allow us to generate non-linear optics near the single photon level \citep{30,31,32} and have plausible applications in the field of quantum networking and quantum information plateforms \citep{33,34}. The necessary condition to realize such potential applications is complete coherence of the quantum device. An all optical switch is one such quantum device with single QD strongly coupled to a nanocavity \citep{35,36,37}. 
The capability of a QD in a micro-cavity producing non-linear optical effects inspire us to study the possibility of manipulating the photon statistics of these systems for practical applications. Due to this, the effects of an optical Kerr medium \citep{39}, an optical parametric amplifier (OPA) \citep{40,41}, or both these nonlinear media \citep{42} on the photon statistics and stability limit of the system have been investigated earlier. It has been demonstrated \citep{39} that due to photon blockade induced by Kerr medium inside a cavity, normal mode splitting (NMS) weakens appreciably.

In the field of quantum theory, entanglement plays a vital role as well as acts as a basic resource in quantum information processing which is useful in tasks of communication and computation \citep{43}. In a double-cavity system entanglement has been observed between different optical modes and mechanical mode of the resonator  \citep{45,46,47,48}. Entangled optomechanical systems \citep{49,50} have the capability to realize quantum communication networks where mechanical modes act as local nodes to store and retrieve the quantum information and optical modes are used to carry the information between these nodes \citep{51,52,53,54}. Therefore such hybrid systems are useful in quantum telecloning \citep{52}, entanglement swapping \citep{54} and quantum teleporation \citep{51,53}.

By combining the tools of cavity electrodynamics with those of quantum dots and Kerr medium, we investigate a novel system consisting of a double semiconductor microcavity with one of the cavity having quantum dot embedded inside it and the other cavity having one set of movable distributed Bragg reflectors (DBR) and an embedded Kerr substrate. In particular, we study the optical bistability, NMS and entanglement properties of this hybrid system. The optical bistability is analyzed for possible signatures of all optical switching. The displacement spectrum of the movable DBR is analyzed to study the NMS. Finally we also focus on the possibility of coherently controlling the entanglement between the various modes of the system.

\section{The Basic Theoretical Model}

The system considered here is a hybrid optomechanical system consisting of two optically coupled semiconductor microcavities A and B supporting two field modes as shown in Fig.1A. Here, cavity A is an optomechanical cavity with one mirror movable and cavity B confines a two-level quantum dot (QD). In addition to this, cavity A has non-linear Kerr medium. The photons are injected into the left cavity A through an external pump with frequency $\omega_{L}$ which exerts radiation pressure on the movable end mirror.  Photons are able to tunnel between these two cavities. Thus, the output optical field from the left cavity A drives the right cavity B. Hence, the confined QD couples to the field mode of the micro-cavity B.  The proposed hybrid system has two distinct non-linearities. First, the Kerr-nonlinear substrate introduces the optical non-linearity. Second, the system has an optomechanical non-linearity due to the coupling of the single optical mode in the micro-cavity A to the micro-mechanical resonator.

The two optically coupled cavities are fabricated with the help of a set of distributed Bragg reflectors (DBR). Light confinement is achieved by the combined action of DBR along the x-direction and air guiding dielectric which provides confinement in the y-z plane \citep{55}. DBR mirror consists of quarter-wavelength thick high and low refractive index layers. The reflectance of DBR is proportional to the number of pairs and the difference between high and low index pairs \citep{56}. The first and the last layers are AlGaAs. This enhances the coupling of light in/out of the structure since the refractive index of AlGaAs lies between those of GaAs and air \citep{56}. GaAs based mechanical resonators are fabricated by utilizing standard micromachining techniques with selective etching \citep{57,58}. A Kerr nonlinear substrate can be deposited on the GaAs cavity according to known experimental technique \citep{59}.

The proposed theoretical model can be described by the optomechanical Hamiltonian in rotating-wave and dipole approximation as

\begin{eqnarray}\label{hom}
H &=& \frac{\hbar\omega_{m}}{2}(p^{2}+q^{2})+\hbar\Delta_{a}a^{\dagger}a+\hbar\Delta_{b}b^{\dagger}b+\frac{\hbar}{2}\Delta_{d}\sigma_{z}\nonumber \\
&+&\hbar g(b\sigma^{+}+b^{\dagger}\sigma^{-})+\hbar J(b^{\dagger}a+b a^{\dagger})-\hbar g_{om} a^{\dagger} a q+\frac{\hbar}{2}\eta a^{\dagger2}a^{2}+i\hbar\epsilon(a^{\dagger}-a);
\end{eqnarray}

\begin{figure}[ht]
\hspace{-0.0cm}
\begin{tabular}{cc}
\includegraphics [scale=0.55]{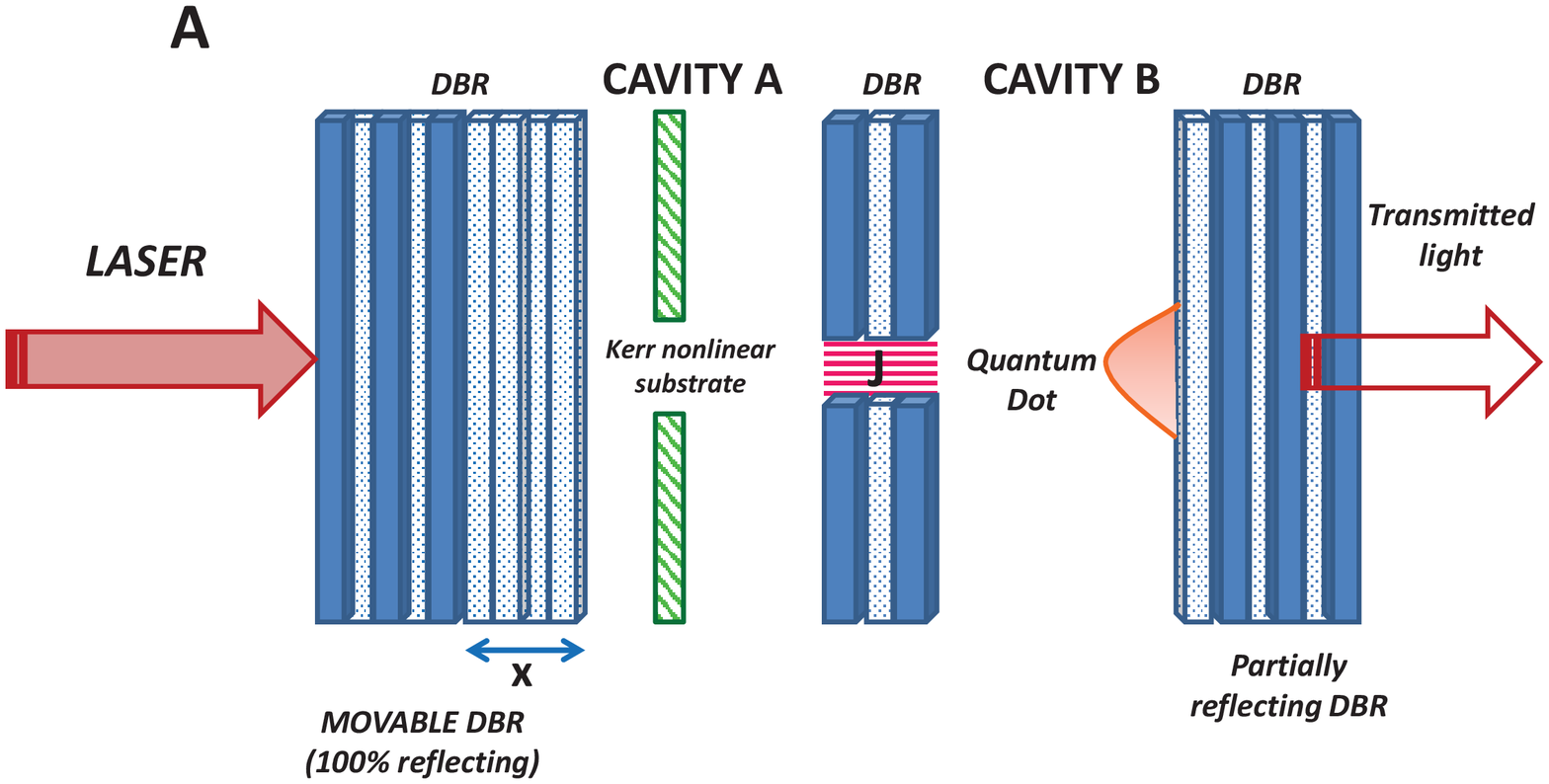}\\
\includegraphics [scale=0.6] {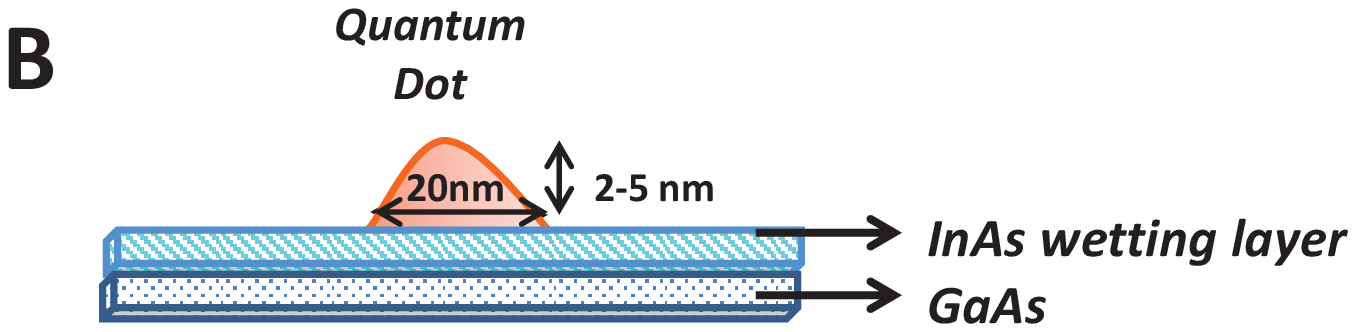}\\
 \end{tabular}
\caption{(Color online) Schematic representation of the setup studied in the text. \textbf{A}: Schematic view of two semiconductor micro-cavities made up of DBR mirrors is shown. Micro-cavity A has a nonlinear Kerr substrate and is driven by a strong pump laser. The left side DBR of micro-cavity A is movable and interacts with the cavity mode via radiation pressure. The right side micro-cavity labelled B confines a quantum dot which is coupled to the field mode.  The two micro-cavities are optically coupled via the tunneling of photons between the two cavitiies. The blue and white strips correspond to AlGaAs and GaAs layers respectively.  \textbf{B}: Figure displays InAs quantum dot embedded inside the array of GaAs layers.}
\label{modfig}
\end{figure}

Here, the first term describes the energy of mechanical oscillator (movable DBR), where $p$ and $q$ are the normalized momentum and position operators respectively. The second and the third terms denote the free energies of the optical modes of cavity A and B respectively with $a$  ($a^\dagger$) and $b$ ($b^\dagger$) as the annihilation (creation) operators respectively. Here, $\Delta_{a}=\omega_{a}-\omega_{L}$ and $\Delta_{b}=\omega_{b}-\omega_{L}$ represents the detuning of cavity A and B respectively with respect to the pump laser. Also $\omega_{a}$ and $\omega_{b}$ are the corresponding cavity resonance frequencies. The fourth term represent the energy of the two-level semiconductor quantum dot with $\Delta_{d}=\omega_{d}-\omega_{L}$ as the detuning of quantum dot. Here, $\omega_{d}$ is the transition frequency between two levels of the quantum dot. Also, $\sigma^{+}$ and $\sigma^{-}$ are the raising and lowering operators respectively of the two-level quantum dot. The fifth term represents the coupling between the QD and optical mode of cavity B with $g$ as the coupling parameter. The sixth term describes the tunneling of photons between the two micro-cavities with $J$ being the tunneling constant. The seventh term represents the coupling between mechanical DBR and the optical mode of cavity A with mirror-photon coupling strength $g_{om}=\frac{\omega_{a}}{L}\sqrt{\frac{\hbar}{m\omega_{m}}}$ with $L$ as the length of the cavity. The eighth term describes the Kerr non-linearity with $\eta=3\omega_{a}^{2}Re[\chi^{(3)}/2\epsilon_{0}V_{c}]$ as the anharmonicity parameter proportional to the third order nonlinear susceptibility $\chi^{(3)}$ of the Kerr medium inside the cavity \citep{60} and $V_{c}$ being the volume of the cavity. The last term describes the external pump of amplitude $\epsilon$ which drives the cavity.

Using the Hamiltonian given by Eqn. (\ref{hom}), the dynamics of the system can be described by the following set of quantum Langevin equations

\begin{equation}\label{1a}
\dot{a}(t)=-i \Delta_{a} a(t)-i\eta a^{\dagger}(t)a^{2}(t)-iJb(t)+\epsilon+ig_{om}a(t)q(t)-k_{a}a(t)+\sqrt{k_{a}}a_{in}(t),
\end{equation}

\begin{equation}\label{1b}
\dot{b}(t)=-i\Delta_{b}b(t)-ig\sigma^{-}(t)-iJa(t)-k_{b}b(t)+\sqrt{k_{b}}b_{in}(t),
\end{equation}

\begin{equation}\label{1c}
\dot{\sigma}(t)=-i\Delta_{d}\sigma^{-}(t)+igb(t)\sigma_{z}(t)-k_{d}\sigma^{-}(t),
\end{equation}

\begin{equation}\label{1d}
\dot{q}(t)=\omega_{m}p(t),
\end{equation}

\begin{equation}\label{1e}
\dot{p}(t)=-\omega_{m}q(t)+g_{om}a^{\dagger}(t)a(t)-\gamma_{m}p(t)+\zeta(t).
\end{equation}

The system interacts with external degrees of freedom that give rise to dissipation. Therefore, $k_{a}$ and $k_{b}$ are introduced as decay constants for the fields in cavity A and B respectively. Here, $a_{in}(t)$ and $b_{in}(t)$ represents input vacuum noise operator with zero mean value for the fields of cavity A and B respectively. Further, $k_{d}$ represents the spontaneous emission decay rate of quantum dot. The mechanical DBR which couples to the mode of cavity A is damped with decay constant $\gamma_{m}$. Equivalently a thermal bath at a damping rate $\gamma_{m}$ is connected to the DBR with a bath occupation given as

\begin{equation}
n_{th}=[\exp{\left(\frac{\hbar\omega_{m}}{k_{B}T}\right)}-1]^{-1},
\end{equation}

 where $k_{B}$ is Boltzmann constant and $T$ is the temperature of the mechanical bath. Moreover, a Brownian force which is described by the operator $\zeta(t)$ corresponds to the noise of the mechanical mode. The correlation functions for all input noise operators are explicitly given in Appendix A.

We are now interested in finding the steady state solutions of the Eqns.(2)-(6). To this end, we first replace the operators by their corresponding mean classical values.  The equations thus obtained are then solved by equating the time derivatives to zero. This yields,

\begin{equation}\label{2a}
\sigma^{-}_{s}=\frac{igb_{s}<\sigma_{z}>_{s}}{k_{d}+i\Delta_{d}},
\end{equation}

\begin{equation}\label{2b}
b_{s}=\frac{-iJa_{s}}{k_{b}+i\Delta_{b}-\frac{g^2<\sigma_{z}>_{s}}{k_{d}+i\Delta_{d}}},
\end{equation}

\begin{equation}\label{2c}
p_s=0,
\end{equation}

\begin{equation}\label{2d}
q_{s}=\chi|a_s|^{2},
\end{equation}

\begin{equation}\label{2e}
a_s=\frac{\epsilon}{(k_a+i\Delta)+\frac{J^2}{k_b+i\Delta_b-\frac{g^{2}<\sigma_{z}>_{s}}{k_{d}+i\Delta_{d}}}}.
\end{equation}

In the above equations, $\chi=\frac{g_{om}}{\omega_{m}}$ is the rescaled optomechanical coupling constant. Also, $\Delta=\Delta_{a}+\eta|a_s|^2-\omega_{m}\chi^{2}|a_s|^2$ is the effective detuning of cavity A.

We now proceed to study the dynamics of quantum fluctuations of the system around its steady state. Therefore, the quantum Langevin equations are linearized around their steady state values as $a(t)\rightarrow a_s+a(t)$, $b(t)\rightarrow b_s+b(t)$, $q(t)\rightarrow q_s+q(t)$ and $p(t)\rightarrow p_s+p(t)$. Assuming that the operators describing the QD has zero fluctuations, we have $\sigma(t)\rightarrow \sigma_s$. We also introduce the amplitude and phase quadratures for the fields as $x_{1}=(b+b^{\dagger})/\sqrt{2}$, $y_{1}=i(b^{\dagger}-b)/\sqrt{2}$, $x_{2}=(a+a^{\dagger})/\sqrt{2}$, $y_{2}=i(a^{\dagger}-a)/\sqrt{2}$, $x_{1in}=(b_{in}+b_{in}^{\dagger})/\sqrt{2}$, $y_{1in}=i(b_{in}^{\dagger}-b_{in})/\sqrt{2}$, $x_{2in}=(a_{in}+a_{in}^{\dagger})/\sqrt{2}$ and $y_{2in}=i(a_{in}^{\dagger}-a_{in})/\sqrt{2}$. The corresponding quantum Langevin equations for the quadratures are:

\begin{equation}\label{3a}
\dot{x_{1}}(t)=\Delta_{b}y_{1}(t)+Jy_{2}(t)-k_{b}x_{1}(t)+\sqrt{k_{b}}x_{1in}(t),
\end{equation}

\begin{equation}\label{3b}
\dot{y_{1}}(t)=-\Delta_{b}x_{1}(t)-Jx_{2}(t)-k_{b}y_{1}(t)+\sqrt{k_{b}}y_{1in}(t),
\end{equation}

\begin{equation}\label{3c}
\dot{x_{2}}(t)= \Delta_1 y_2(t)+Jy_1(t)+\Gamma_{1}x_{2}(t)+\delta_{1}y_{2}(t)+ia_{-}q(t)-k_{a}x_{2}(t)+\sqrt{k_{a}}x_{2in}(t),
\end{equation}

\begin{equation}\label{3d}
\dot{y_{2}}(t)=-\Delta_1 x_2(t)-Jx_1(t)-\Gamma_{1}y_{2}(t)+\delta_{1}x_{2}(t)+a_{+}q(t)-k_{a}y_{2}(t)+\sqrt{k_{a}}y_{2in}(t),
\end{equation}

\begin{equation}\label{3e}
\dot{q}(t)=\omega_{m}p(t),
\end{equation}

\begin{equation}\label{3f}
\dot{p}(t)=-\omega_{m}q(t)+a_{+}x_{2}(t)-ia_{-}y_{2}(t)-\gamma_{m}p(t)+\zeta(t),
\end{equation}

where,

\begin{equation}\label{4a}
\Gamma_{1}=\frac{-i\eta(a_{s}^{2}-a_{s}^{*2})}{2},
\end{equation}

\begin{equation}\label{4b}
\delta_{1}=\frac{-\eta(a_{s}^{2}+a_{s}^{*2})}{2},
\end{equation}

\begin{equation}\label{4c}
\Delta_{1}=\Delta+\eta|a_s|^2,
\end{equation}

\begin{equation}\label{4d}
a_{\pm}=\omega_{m}\chi\frac{(a_{s}\pm a_{s}^{*})}{\sqrt{2}}.
\end{equation}

\section{Optical Bistability}

In various non-linear systems, bistability is a pervasive phenomenon. The phenomena of optical bistability has been demonstrated in many of the optomechanical systems \citep{61,62,63,64,65,66,67,68,69}. All these systems are characterized by a high degree of nonlinearity that arises because of dynamical back action induced by radiation pressure. These kind of optomechanical systems have potential applications in memory storage \citep{70,71} and all optical switching devices \citep{72,73}.

As the first insight, we will discuss how the different physical parameters of the system affect the optical bistability. In the present system, the optomechanical coupling and Kerr nonlinearity introduces two distinct nonlinearities. From the steady state solutions (\ref{2a})-(\ref{2e}), the steady state expression for $a_{s}$ is written as

\begin{equation}\label{5c}
a_{s}=\frac{\epsilon}{k_{a}+i\Delta'+i\eta|a_{s}|^2+\frac{J^2}{k_{b}+i\Delta_{b}-\frac{g^2<\sigma_{z}>_s}{k_{d}+i\Delta_{d}}}},
\end{equation}

where, $\Delta'=\Delta_{a}-\omega_{m}\chi^2|a_{s}|^2$. The simplified equation that indicates the existence of bistable behavior of the optical field derived from eqn. (\ref{5c}) is given as

\begin{equation}\label{6a}
|a_{s}|^2[k_{n}^2+(\Delta_{n}+(\eta-\omega_{m}\chi^2)|a_{s}|^2)^2]=\epsilon^2,
\end{equation}

where, $k_{n}=k_{a}+\frac{J^2(k_{d}A_{1}+\Delta_{d}A_2)}{A_{1}^2+A_{2}^2}$, $\Delta_{n}=\Delta_{a}+\frac{J^2(\Delta_{d}A_{1}-k_{d}A_{2})}{A_{1}^2+A_{2}^2}$, $A_{1}=k_{b}k_{d}-\Delta_{b}\Delta_{d}-g^2<\sigma_{z}>_{s}$ and $A_{2}=\Delta_{b}k_{d}+k_{b}\Delta_{d}$. Here, $k_{n}$ and $\Delta_{n}$ represent effective decay rate and detuning of cavity A in presence of quantum dot.

\begin{figure}[ht]
\hspace{-0.0cm}
\includegraphics [scale=1.0] {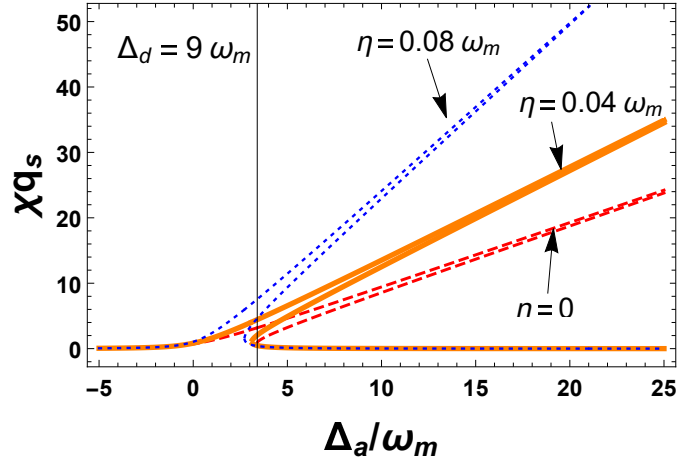}
\caption{(Color online) Plots of $\chi q_{s}$ versus $\frac{\Delta_{a}}{\omega_{m}}$ for three different values of Kerr-nonlinear parameter ($\eta=0$ (dashed line), $\eta=0.04\omega_{m}$ (solid line), $\eta=0.08\omega_{m}$ (dotted line)) in high power limit $<\sigma_{z}>_{s}=0$. Other parameters used are $k_{a}=0.1\omega_{m}$, $k_{d}=1.8\omega_{m}$,  $J=\omega_{m}$, $k_{b}=0.1\omega_{m}$, $g=0.5\omega_{m}$, $\chi=0.3\omega_{m}$, $\Delta_{b}=\omega_{m}$ and $\epsilon=4\omega_{m}$.}
\label{ob1}
\end{figure}

\begin{figure}[ht]
\hspace{-0.0cm}
\begin{tabular}{cc}
\includegraphics [scale=0.9]{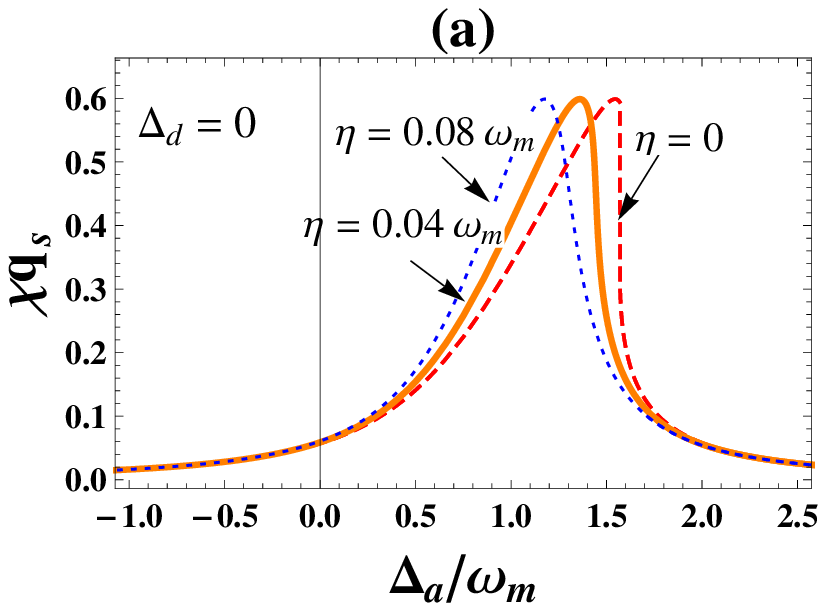}& \includegraphics [scale=0.9] {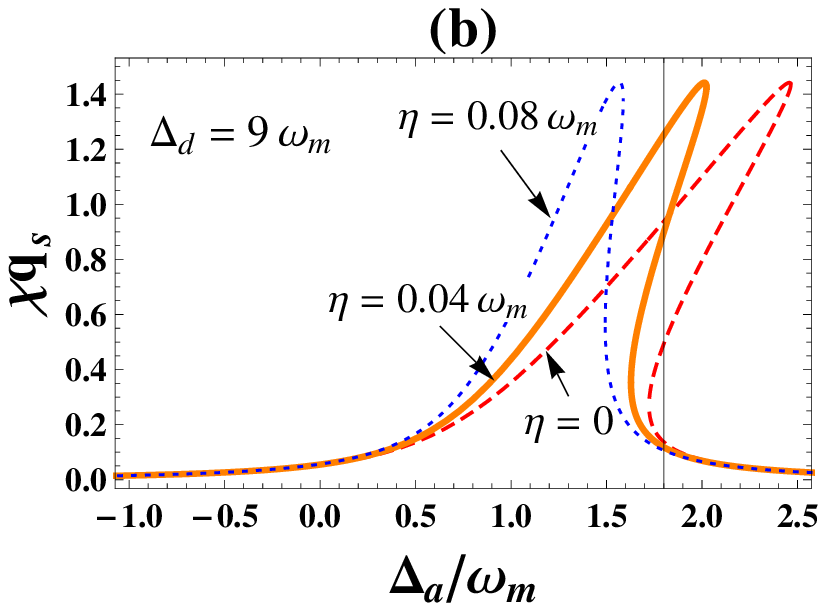}\\
 \end{tabular}
\caption{(Color online) Plots of $\chi q_{s}$ versus $\frac{\Delta_{a}}{\omega_{m}}$ for three different values of Kerr-nonlinear parameter ($\eta=0$ (dashed line), $\eta=0.04\omega_{m}$ (solid line), $\eta=0.08\omega_{m}$ (dotted line)) in low power limit $<\sigma_{z}>_{s}=-1$ for $\Delta_{d}=(0)$ (plot a) and $\Delta_{d}=9 \omega_{m}$ (plot b). Here, $\epsilon=0.7\omega_{m}$ and other parameters are same as in fig.(\ref{ob1}).}
\label{ob2}
\end{figure}

The bistable behavior of the system is analyzed in two different limits. The first limit is the high power limit where the steady state value of $<\sigma_{z}>_{s}=0$ i.e. both excited and ground states of the QD are equally populated. The second limit is the low power limit in which the steady state value $<\sigma_{z}>_{s}=-1$ i.e. only ground state of QD is populated. Figure \ref{ob1} shows the bistable behavior in the high power limit ($<\sigma_{z}>_{s}=0$). The figure displays the $\chi q_{s}$ ( proportional to intracavity optical intensity as shown in Eq.(11)) with respect to detuning of micro-cavity A in the high power limit for different values of Kerr-nonlinear parameter ($\eta=0$ (dashed line), $\eta=0.04\omega_{m}$ (solid line), $\eta=0.08\omega_{m}$ (dotted line)) at $\Delta_{d}=5k_{d}$. It is clear from the figure that with change in Kerr-nonlinearity parameter, bistability of the system changes. The bistable behavior of the intracavity photons does not change with Kerr-nonlinear parameter in the absence of QD detuning (figure not shown). Figures \ref{ob2}(a) and \ref{ob2}(b) show the bistable behavior in the low power limit ($<\sigma_{z}>_{s}=-1$) for $\Delta_{d}=0$ and $\Delta_{d}=5$ respectively for three different values of Kerr-nonlinear parameter. As the two plots in figure \ref{ob2} are compared, it is observed that bistability is dependent on QD detuning and the bistable behavior increases with increase in QD detuning. Figure \ref{ob2}(b) displays that the cavity A detuning range for the occurrence of bistable behavior is significantly different compared to  Fig. 3(a) for the same input laser power as the Kerr-nonlinearity is varied. Further, comparing the plots in Fig. \ref{ob1} and \ref{ob2}, the bistable behavior is seen to be enhanced in the high power limit. Also in the high power limit, bistability occurs at high detuning value of cavity A. It should be noted that onset of bistable behavior is due to the two inherent nonlinearities in the system (Kerr-nonlinearity and optomechanical coupling).

\begin{figure}[ht]
\hspace{-0.0cm}
\begin{tabular}{cc}
\includegraphics [scale=0.9]{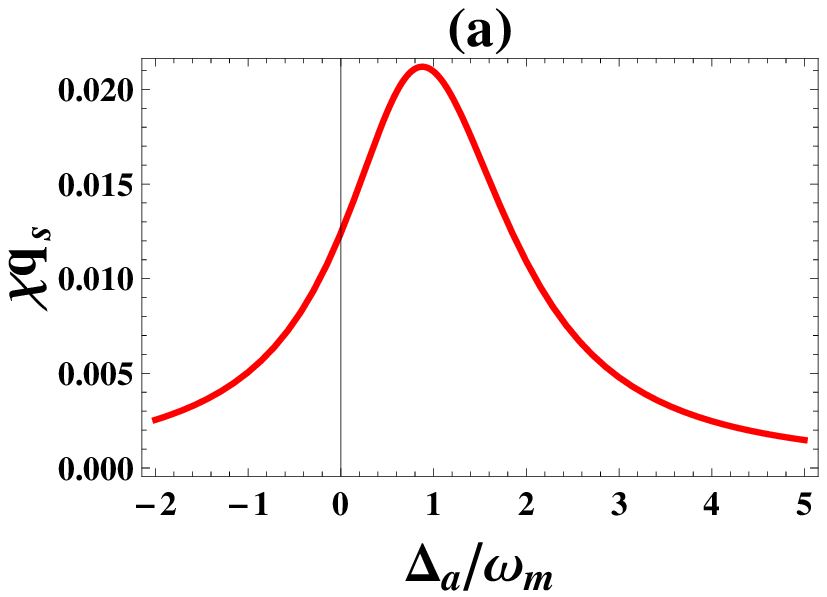}& \includegraphics [scale=0.9] {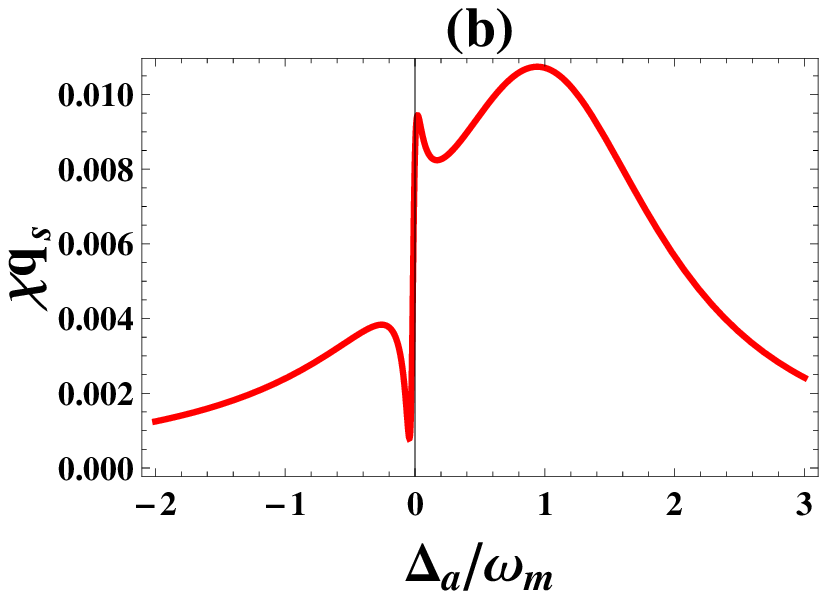}\\
 \end{tabular}
\caption{(Color online) Plots of $\chi q_{s}$ versus $\frac{\Delta_{a}}{\omega_{m}}$. Plot (a) is for high power limit $<\sigma_{z}>_{s}=0$ and $\epsilon=4\omega_{m}$ while plot (b) is for low power limit $<\sigma_{z}>_{s}=-1$ and $\epsilon=0.4\omega_{m}$. Other system parameters used are $k_{a}=\omega_{m}$, $k_{d}=0.01\omega_{m}$, $\Delta_{b}=\omega_{m}$, $J=\omega_{m}$, $k_{b}=0.1\omega_{m}$, $g=0.2\omega_{m}$, $\Delta_{d}=-\Delta_{a}$, $\eta=0.01\omega_{m}$ and $\chi=0.04\omega_{m}$.}
\label{ob3}
\end{figure}

Figures \ref{ob3} illustrates the asymmetric switching behavior in the two limits ($<\sigma_{z}>_{s}=0$ and $<\sigma_{z}>_{s}=-1$) when QD detuning and cavity detuning are same ($\Delta_{d}=-\Delta_{a}$) at Kerr-nonlinearity parameter $\eta=0.01\omega_{m}$. By comparing the two plots in figure \ref{ob3}, we observe that characteristics of all-optical switch is exhibited when $<\sigma_{z}>_{s}=-1$. This asymmetric nature of split resonance around $\Delta_{a}=0$ (fig. \ref{ob3}(b)) is due to the presence of Kerr and optomechanical nonlinearity. However in the limit when $<\sigma_{z}>_{s}=0$ , the system fails to display the optical switching characteristic as shown in fig. \ref{ob3}(a).

Conclusively, the Kerr and optomechanical nonlinearities are the cause of bistability in the system. The external parameters i.e. Kerr nonlinearity ($\eta$), optomechanical coupling ($\chi$), power of driving laser ($\epsilon$), cavity-field detuning of optomechanical cavity ($\Delta_{a}$) and QD-field detuning ($\Delta_{d}$) controls the switching behaviour of the intracavity optical intensity in the optomechanical cavity between its two stable branches. Moreover, in the high power limit when the two states of QD are equally populated ($<\sigma_{z}>_{s}=0$), bistability is absent and the system fails to operate as an all-optical switch. However, in the low power limit of $<\sigma_{z}>_{s}=-1$, i.e. only ground state of QD is populated, the system demonstrates the behavior of an all-optical switch.

\section{Normal Mode Splitting}

In an optomechanical cavity, one of the vital phenomena observed in the displacement spectrum of the movable cavity mirror is Normal Mode Splitting (NMS) which is due to mixing of the fluctuations of different modes in the system around their steady state \citep{74,75,76}. Due to the strong coupling between different modes of the system, there is energy exchange among them on a time scale faster than decoherence time of each mode due to which NMS occurs. In various optomechanical experiments, it leads to ground state cooling of the mechanical oscillator \citep{74,76,77}.

In this section, we study the normal mode splitting (NMS) and evaluate the spectrum of small fluctuations in the position quadrature of the movable DBR in the presence of quantum dot and Kerr non-linearity. In order to calculate the NMS, the equations of motion in time domain (eqns. (\ref{3a})-(\ref{3f})) are transformed into frequency domain using Fourier transform. The Fourier transformed equations are then solved for the displacement spectrum. In Fourier space, the displacement spectrum is defined as,

\begin{equation}\label{7a}
S_q(\omega)=\frac{1}{4\pi}\int{d\omega'e^{-i(\omega+\omega')t}<\delta q(\omega)\delta q(\omega')+\delta q(\omega')\delta q(\omega)>}.
\end{equation}

In the steady state, the system is always stable. Hence, the stability conditions given in Appendix B are satisfied by the system. Thus, using the correlation functions in Fourier space given in Appendix A (42-50), we obtain the displacement spectrum of the oscillating DBR as

\begin{equation}\label{7b}
S_{q}(\omega)=\frac{1}{X_{1}(\omega)X_{1}(-\omega)}[X_{2}(\omega)X_{2}(-\omega)+X_{3}(\omega)X_{3}(-\omega)+X_{4}(\omega)X_{4}(-\omega)+X_{5}(\omega)X_{5}(-\omega)+X_{6}(\omega)].
\end{equation}

The coefficients $X_{i}( \omega)$ (i=1,2,3,4,5,6) appearing in the expression Eqn.(28) are given in Appendix C.

\begin{figure}[ht]
\hspace{-0.0cm}
\begin{tabular}{cc}
\includegraphics [scale=0.9]{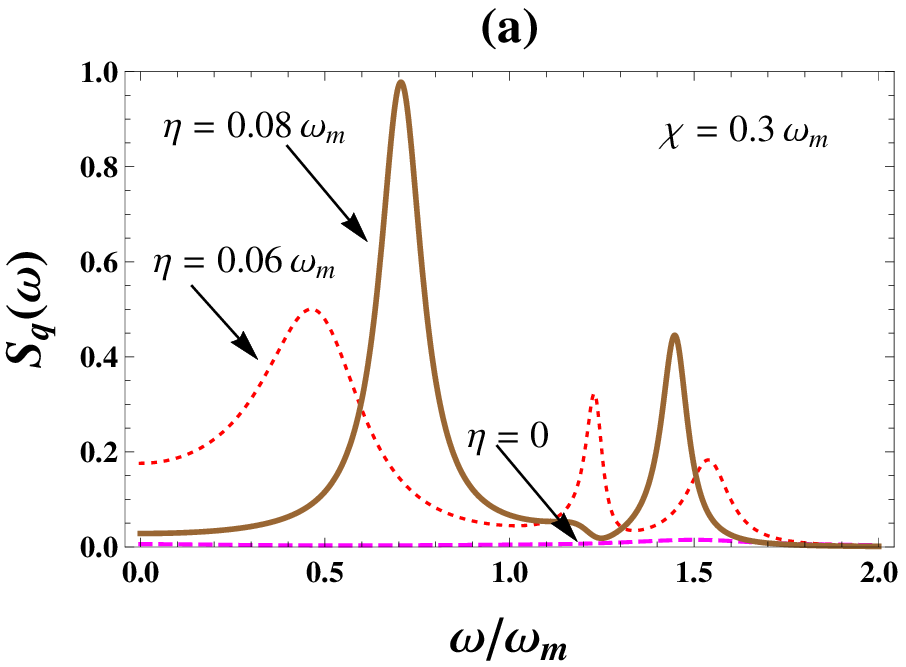}& \includegraphics [scale=0.9] {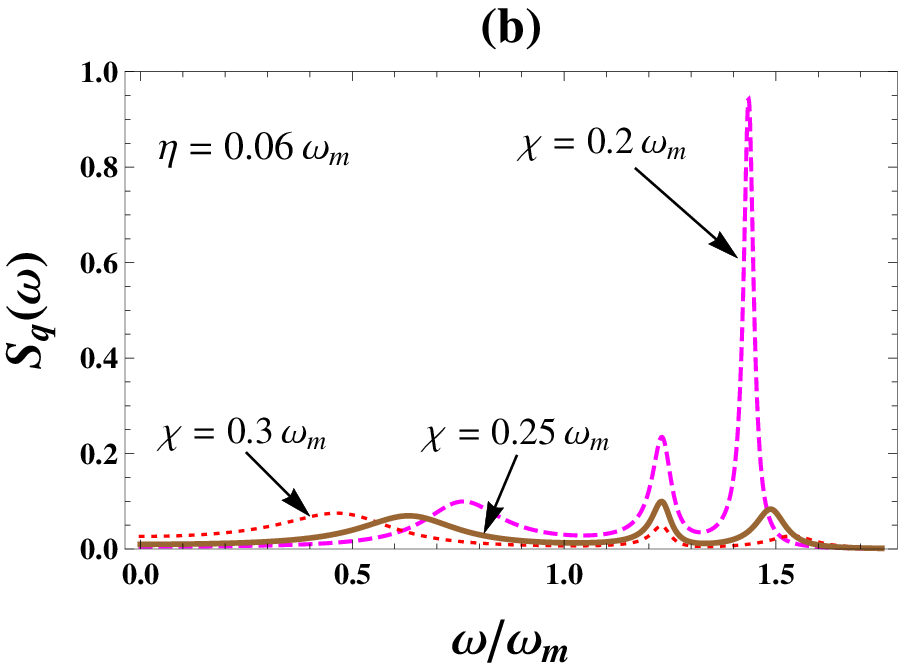}\\
 \end{tabular}
\caption{(Color online) Displacement Spectrum as a function of dimensionless frequency for high power limit $<\sigma_{z}>_{s}=0$. In above figures, plot (a) shows the variation for different values of Kerr nonlinearity parameter ($\eta$) and plot (b) shows the variation for various values of optomechanical coupling parameter ($\chi$). Here, $k_{B}T/\hbar\omega_{m}=10^{6}$, $k_{a}=0.5\omega_{m}$, $\Delta=-0.1\omega_{m}$, $\gamma_{m}=10^{-5}\omega_{m}$, $k_{d}=1.8\omega_{m}$, $\Delta_{b}=-1.2\omega_{m}$, $J=\omega_{m}$, $k_{b}=0.15\omega_{m}$, $\Delta_{d}=5k_{d}$, $g=0.5\omega_{m}$, $\epsilon=8\omega_{m}$ and $\eta=0.06\omega_{m}$.}
\label{ds1}
\end{figure}

Figure \ref{ds1}(a) illustrates displacement spectrum ($S_{q}(\omega)$) of the movable DBR as a function of dimensionless frequency ($\omega/\omega_{m}$). In the high power limit $<\sigma_{z}>_{s}=0$, the displacement spectrum is observed for three different values of Kerr non-linear parameter, $\eta=0$ (dashed line), $\eta=0.06\omega_{m}$ (dotted line), $\eta=0.08\omega_{m}$ (solid line) at optomechanical coupling $\chi=0.3\omega_{m}$. Figure \ref{ds1}(b) shows the displacement spectrum ($S_{q}(\omega)$) for the mechanical DBR versus the dimensionless frequency ($\omega/\omega_{m}$) for various values of optomechanical coupling, $\chi=0.2\omega_{m}$ (dashed line), $\chi=0.25\omega_{m}$ (solid line), $\chi=0.3\omega_{m}$ (dotted line) at Kerr-nonlinearity factor $\eta=0.06\omega_{m}$ in the high power limit $<\sigma_{z}>_{s}=0$. In both the plots, the NMS displays three peaks. This NMS is due to the coupling between the fluctuations of the mechanical mode of movable DBR and fluctuations of the two optical modes. There is a coherent exchange of energy between the three modes. To observe the phenomena of NMS, it is important that decoherence of each mode should be less than the timescale for exchange of energy between the three modes. Figure \ref{ds1}(a) shows that in the absence of Kerr-nonlinearity ($\eta=0$), only one peak with extremely small amplitude is visible. From fig. \ref{ds1}(b), it is observed that when the optomechanical coupling is $\chi=0.2\omega_{m}$,  then one of the peak has very high amplitude. This observation is because of the fact that when optomechanical coupling is less there is less energy exchange between the mechanical and the optical mode of cavity A. Thus the energy exchange between the two optical modes becomes dominant which is observed as the high amplitude peak.

\begin{figure}[ht]
\hspace{-0.0cm}
\begin{tabular}{cc}
\includegraphics [scale=0.9]{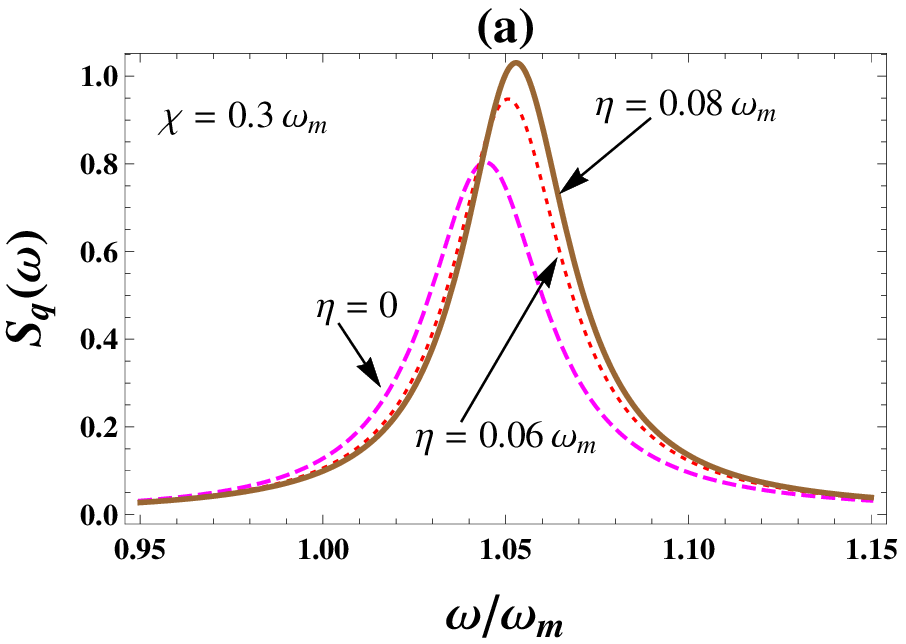}& \includegraphics [scale=0.9] {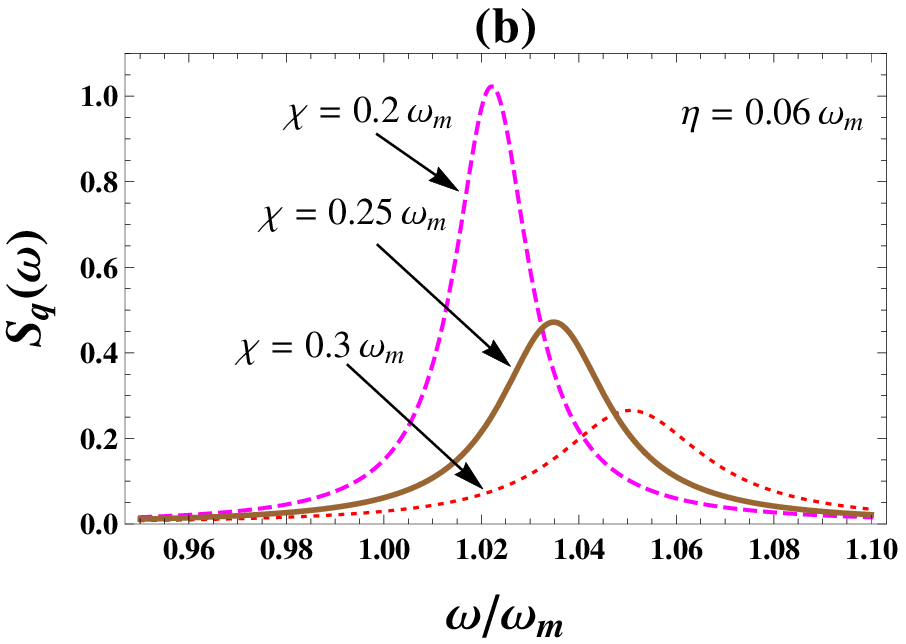}\\
 \end{tabular}
\caption{(Color online) Displacement Spectrum as a function of dimensionless frequency for low power limit $<\sigma_{z}>_{s}=-1$. In above figures, plot (a) shows the variation for different values of Kerr nonlinearity parameter ($\eta$) and plot (b) shows the variation for various values of optomechanical Coupling parameter ($\chi$). Here, $\epsilon=0.8\omega_{m}$ , $J=0.7\omega_{m}$ and rest of the parameters are same as in fig. (\ref{ds1}).}
\label{ds2}
\end{figure}

In fig. \ref{ds2}(a)  the displacement spectrum  of the movable DBR as a function of dimensionless frequency ($\omega/\omega_{m}$) is plotted for three different values of Kerr non-linear parameter, $\eta=0$ (dashed line), $\eta=0.06\omega_{m}$ (dotted line), $\eta=0.08\omega_{m}$ (solid line) at optomechanical coupling $\chi=0.3\omega_{m}$ in the low power limit $<\sigma_{z}>_{s}=-1$. Figure \ref{ds2}(b) represents the displacement spectrum for various values of optomechanical coupling, $\chi=0.2\omega_{m}$ (dashed line), $\chi=0.25\omega_{m}$ (solid line), $\chi=0.3\omega_{m}$ (dotted line) and Kerr-nonlinearity factor $\eta=0.06\omega_{m}$ in the low power limit $<\sigma_{z}>_{s}=-1$. As observed, both the plots does not show NMS. This is due to the fact that in the low power limit ($<\sigma_{z}>_{s}=-1$), the energy exchange is not efficient between different modes. Fig. \ref{ds2}(a) also illustrates that with change in value of Kerr-nonlinear parameter, the displacement spectrum does not show much variation. However, the displacement spectrum peak enhances with decrease in the optomechanical coupling parameter as shown in fig. \ref{ds2}(b).

\section{Entanglement}

Entanglement plays the major role in various quantum information applications like quantum computation, quantum metrology and quantum cryptography \citep{78,79,80}. Recently, \citep{81} the control of entanglement dynamics in a system of three coupled quantum oscillators had been shown. Moreover, it is possible to generate entanglement in quantum parametric oscillators using phase control \citep{82}. Also entanglement has been observed in an optomechanical system consisting of quantum well embedded inside  \citep{83} . There are different methods to study the entanglement but one of the commonly used method is using Logarithmic negativity \citep{84}.
In an optomechanical system, stationary entanglement implies strong correlations between phonon and photon mode. In this section, we study numerically the stationary entanglement between different modes of the system. Stationary entanglement is meaningful only when the system is in a single stable state. The system is stable when it satisfies stability conditions obtained using Routh-Hurwitz Criterion given in Appendix B.

Now, in compact matrix form, the system of linearized equations given in Eqns. (\ref{3a})-(\ref{3f}) can be written as

\begin{equation}\label{8a}
\dot{u}(t)=Mu(t)+n(t).
\end{equation}

Here, $u(t)=[q(t),p(t),x_{1}(t), y_{1}(t), x_{2}(t),y_{2}(t)]^T$,

$n(t)=[0,\zeta(t),\sqrt{k_{b}}x_{1in}(t),\sqrt{k_{b}}y_{1in}(t),\sqrt{k_{a}}x_{2in}(t),\sqrt{k_{a}}y_{2in}(t)]^T$ and

\begin{equation}\label{8b}
M = \left[\begin{matrix}
       0 & \omega_{m} & 0 & 0 & 0 & 0\\
       -\omega_{m} & -\gamma_{m} & 0 & 0 & a_{+} & -ia_{-}\\
       0 & 0 & -k_{b} & \Delta_{b} & 0 & J\\
       0 & 0 &-\Delta_{b} & -k_{b} & -J & 0\\
       ia_{-} & 0 & 0 & J & (\Gamma_{1}-k_{a}) & (\Delta_{1}+\delta_{1})\\
      a_{+} & 0 & -J & 0 & (-\Delta_{1}+\delta_{1}) & (-\Gamma_{1}-k_{a})
     \end{matrix}\right].
\end{equation}

\begin{figure}[ht]
\hspace{-0.0cm}
\begin{tabular}{cc}
\includegraphics [scale=0.9]{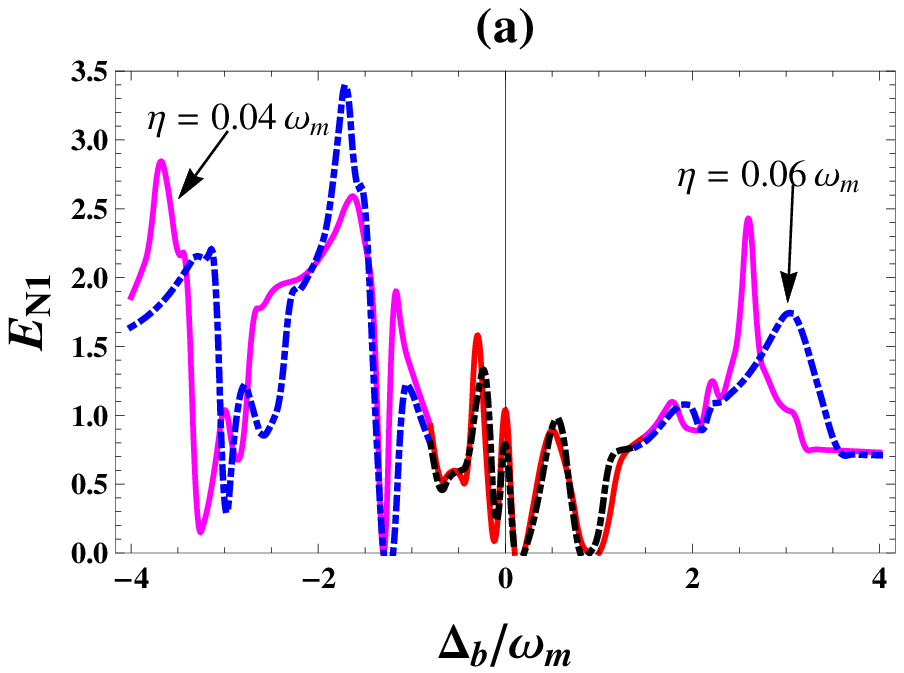}& \includegraphics [scale=0.9] {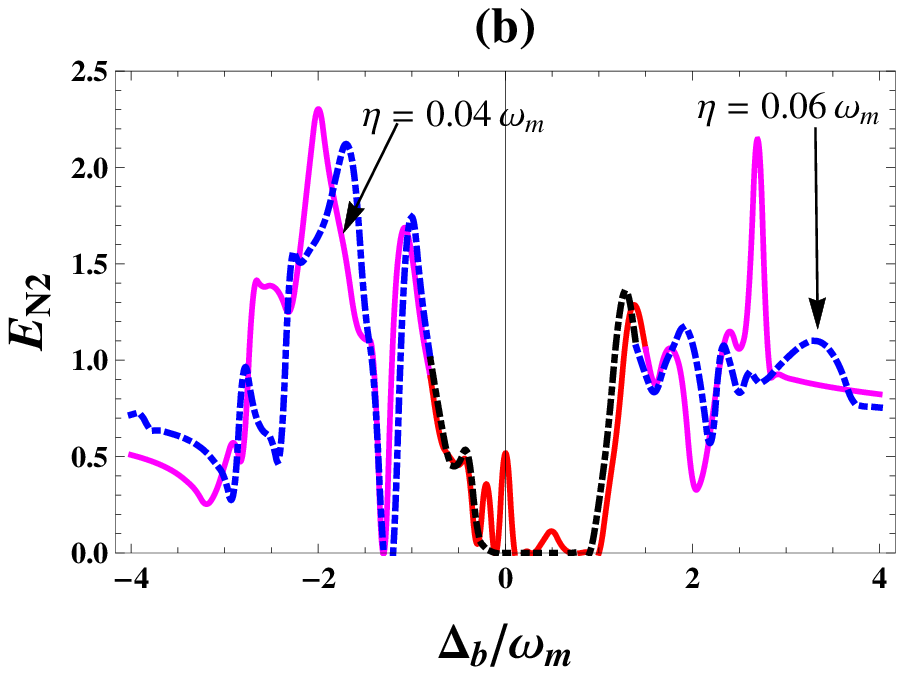}\\
 \end{tabular}
\caption{(Color online) The logarithmic negativity ($E_{N1}$, $E_{N2}$) versus the normalized detuning $\Delta_{b}/\omega_{m}$ for two different values of Kerr nonlinearity parameter ($\eta=0.04\omega_{m}$ (solid line), $\eta=0.06\omega_{m}$ (dot-dashed line)) in the high power limit $<\sigma_{z}>_{s}=0$. Plot (a) shows logarithmic negativity between the mechanical mode and optical mode of cavity B. Plot (b) shows logarithmic negativity between the two optical modes. Here the parameters used are $k_{B}T/\hbar\omega_{m}=10^{4}$, $k_{a}=0.5\omega_{m}$, $\Delta=-0.1\omega_{m}$, $\gamma_{m}=10^{-5}\omega_{m}$, $k_{d}=1.8\omega_{m}$, $J=\omega_{m}$, $k_{b}=0.15\omega_{m}$, $\Delta_{d}=5k_{d}$, $g=0.5\omega_{m}$, $\epsilon=2.2\omega_{m}$ and $\chi=0.5\omega_{m}$.}
\label{EN1}
\end{figure}

The solution to Eqn. (\ref{8a}) is given as $u(t)=F(t)u(0)+\int\limits_0^t{dsF(s)n(t-s)}$ with $F(t)=e^{Mt}$. The steady state of the system is achieved only if it is stable. This is possible only if all the eigenvalues of the drift matrix $M$ have negative real parts so that $F(\infty)=0$. This means that the stability conditions given in Appendix B must be satisfied. Since all the noise are Gaussian in nature and the dynamics of the fluctuations of system is linearized, therefore the steady state of the system is a zero-mean Gaussian state. As a result of this, it can be fully characterized by its $6\times6$ correlation matrix (CM) $V$ with matrix elements given as

\begin{eqnarray}\label{8c}
V_{ij}&=&\frac{u_{i}(\infty)u_{j}(\infty)+u_{j}(\infty)u_{i}(\infty)}{2}\nonumber \\
&=&\int\limits_0^\infty{ds\int\limits_0^\infty{ds'F_{ik}(s)F_{jl}(s')D_{kl}(s-s')}},
\end{eqnarray}

where,
\begin{equation}\label{8d}
D_{kl}(s-s')=\frac{n_{k}(\infty)n_{l}(\infty)+n_{l}(\infty)n_{k}(\infty)}{2}=D\delta(s-s')
\end{equation}

is the diffusion matrix with

\begin{equation}\label{8e}
D=diag[0,\gamma_{m}(2n_{th}+1), k_{b}/2, k_{b}/2, k_{a}/2, k_{a}/2].
\end{equation}

\begin{figure}[ht]
\hspace{-0.0cm}
\begin{tabular}{cc}
\includegraphics [scale=0.9]{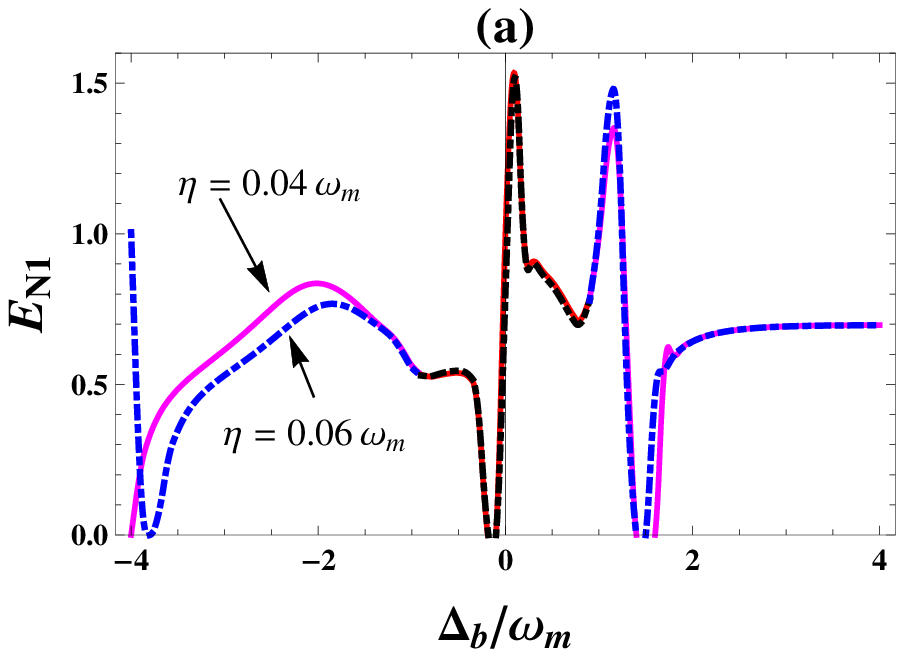}& \includegraphics [scale=0.9] {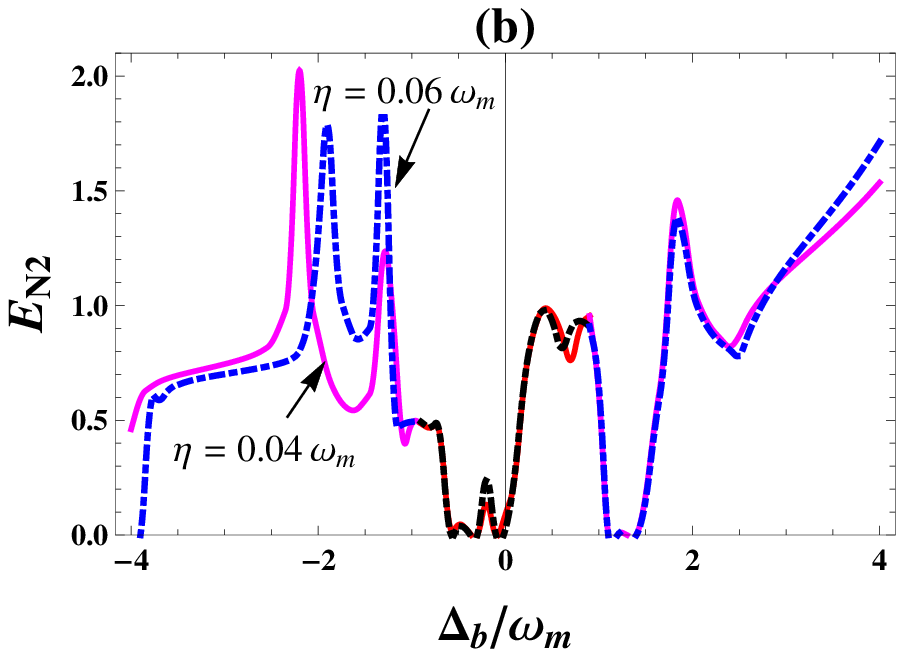}\\
 \end{tabular}
\caption{(Color online) The logarithmic negativity ($E_{N1}$, $E_{N2}$) versus the normalized detuning $\Delta_{b}/\omega_{m}$ for two different values of Kerr nonlinearity parameter ($\eta=0.04\omega_{m}$ (solid line), $\eta=0.06\omega_{m}$ (dot-dashed line)) in the low power limit $<\sigma_{z}>_{s}=-1$. Plot (a) shows logarithmic negativity between the mechanical mode and optical mode of cavity B. Plot (b) shows logarithmic negativity between the two optical modes. Here the parameters used are $J=0.7\omega_{m}$, $k_{b}=0.5\omega_{m}$, $\epsilon=0.8\omega_{m}$ and $\chi=0.3\omega_{m}$. Other parameters are same as in fig.\ref{EN1}.}
\label{EN2}
\end{figure}

It should be noted that to achieve the entanglement between the mechanical and optical modes, it is assumed that the mechanical oscillator should have high mechanical quality factor i.e. $Q=\omega_{m}/\gamma_{m}\rightarrow\infty$ and damping should be weak $\gamma_{m}\rightarrow 0$ so that quantum Brownian noise \citep{85} becomes

\begin{equation}\label{8f}
<\zeta(t)\zeta(t')+\zeta(t')\zeta(t)>\simeq\gamma_{m}(2n_{th}+1)\delta(t-t').
\end{equation}

Hence, Eqn. (\ref{8c}) is rewritten as,

\begin{equation}\label{8g}
V=\int\limits_0^\infty{dsF(s)DF(s)^T}.
\end{equation}

The above equation is equivalent to the following Lyapunov equation \citep{86}

\begin{equation}\label{8h}
MV+VM^T=-D,
\end{equation}

when the stability conditions satisfy $F(\infty)=0$. Eqn. (\ref{8h}) represents a linear matrix equation which can be easily solved for V. However, it is too cumbersome to report here the exact general expression for the same. A trivial measure to compute entanglement $E_{N}$ between any two bipartite subsystem is obtained by tracing out the third mode (i.e. eliminating rows and columns of V that correspond to third mode). The reduced correlation matrix is a $4\times4$ matrix $V'$ which is still a Gaussian one. Therefore, continuous variable (CV) entanglement can be well defined using logarithmic negativity $E_{N}$ as

\begin{equation}\label{8i}
E_{N}=max(0,-2ln2\eta^{-}),
\end{equation}

where $\eta^{-}=\sqrt{(\Sigma(V')-\sqrt{\Sigma(V')^2-4det(V')})}/\sqrt{2}$ is the smallest symplectic eigenvalue of the bipartite system with $\Sigma(V')=det(X)+det(Y)-det(Z)$. Here, $2\times2$ block form of $V'$ can be written as

\begin{equation}\label{8j}
V'=\left[\begin{matrix}
       X & Z\\
       Z^{T} & Y
     \end{matrix}\right].
\end{equation}

As seen in Eqn. (\ref{8i}), the logarithmic negativity is decreasing function of $\eta^{-}$ and it is used to measure the entanglement between two Gaussian states. Entanglement of a Gaussian state is possible only if $\eta^{-}<1/2$ (or $4det(V')<\Sigma(V')-1/4$). This condition is equivalent to Simon's necessary and sufficient condition for entanglement of two Gaussian states \citep{87}.

Figures \ref{EN1} show the stationary entanglement in the two possible bipartitions of the system using logarithmic negativity $E_{N}$. Here, fig. \ref{EN1}(a) represents the logarithmic negativity $E_{N1}$ for the mechanical mode and optical mode of cavity B as a function of cavity B detuning for two different values of Kerr non-linearity parameter ($\eta=0.04\omega_{m}$, $\eta=0.06\omega_{m}$) in the high power limit i.e. $<\sigma_{z}>_{s}=0$. Fig. \ref{EN1}(b) shows the logarithmic negativity $E_{N2}$ for the two optical modes between cavity A and cavity B versus the detuning of cavity B for two different values of Kerr non-linearity parameter ($\eta=0.04\omega_{m}$, $\eta=0.06\omega_{m}$) in the high power limit i.e. $<\sigma_{z}>_{s}=0$. In both the figures, the red ($\eta=0.04\omega_{m}$) and black ($\eta=0.06\omega_{m}$) curve indicate that part of logarithmic negativity, in which the system is not stable since for these values of cavity B detuning ($\Delta_{b}$) stability conditions given in Appendix B are not satisfied. As expected, Kerr-parameter moderates the entanglement in both the cases. This is due to the fact Kerr interaction induces photon blockade. The number of intracavity photon reduces considerably due to this blockade. This effect the optomechanical entanglement as the effective coupling between the mirror and the cavity field due to the radiation pressure effectively reduces. In addition, the coupling between the two optical modes weakens which destroys the entanglement between the two optical modes.

Figures \ref{EN2} display the dependence of steady state entanglement in the low power limit i.e. $<\sigma_{z}>_{s}=-1$. Plot \ref{EN2}(a) shows the logarithmic negativity $E_{N1}$ for the mechanical mode and optical mode of cavity B as a function of cavity B detuning for two different values of Kerr non-linearity parameter ($\eta=0.04\omega_{m}$, $\eta=0.06\omega_{m}$) in the low power limit. Plot \ref{EN2}(b) shows the logarithmic negativity $E_{N2}$ for the two optical modes between cavity A and cavity B versus the detuning of cavity B for two different values of Kerr non-linearity parameter ($\eta=0.04\omega_{m}$, $\eta=0.06\omega_{m}$) in the low power limit. Here also we find that there is no stable state of logarithmic negativities (shown by red and black curve) when $\Delta_{b}/\omega_{m}$ is near the region ($-0.8$ to $1.5$) for both the values of Kerr non-linear parameter ($\eta=0.04\omega_{m}$, $\eta=0.06\omega_{m}$). This can be better understood from eqns.\ref{3a} -\ref{3f}. Due to the photon blockade the entanglement decreases when the Kerr interaction increases. Further, fig. \ref{EN2}(a) shows that the logarithmic negativity $E_{N1}$ for the mechanical mode and optical mode of cavity B is not affected much as the Kerr-nonlinear interaction is changed. This is due the fact that in the low power limit the intensity of pump laser is low which decreases the amount of photons inside the cavity. Hence, the variation of Kerr nonlinear parameter does not affect the optomechanical entanglement much.

All the parameters used in our calculations are accessible in earlier experiments \citep{2,88,89,90,91,92,93,94}. The length of the optical cavity may vary from $10^{-3}-25\times10^{-3} m$. Effective mass of the mechanical mirror can vary between $5-145ng$ and its frequency varies between $1-10MHz$. The corresponding damping rate of the resonator is $\gamma_{m}=\omega_{m}/Q$, where $Q=10^7$ is the Quality factor of the optomechanical cavity. The external laser pump strength can vary from $0.2-0.5\omega_{m}$. Also, the damping rate of the intracavity optical field may vary from $2\pi\times8.75kHz-2\pi\times0.66MHz$ \citep{95,96}. Moreover, the limit $\gamma_{m} \ll \omega_{m} \ll k_{B}T/\hbar$ is always satisfied in typical optomechanical experiments \citep{96,97,98,99}.

\section{Conclusions}

In conclusion, we have analyzed the photon statistics, mechanical displacement spectrum and bipartite entanglement in a hybrid quantum optomechanical system consisting of two optically coupled semiconductor microcavities containing a quantum dot and a Kerr nonlinear substrate. The optical bistability that is generated due to the optical and optomechanical nonlinearity displays a typical optical switching behavior which can be controlled and tuned by appropriately changing the QD-cavity coupling, Kerr nonlinearity, laser power and the optomechanical coupling. The displacement spectrum of the movable DBR exhibits a three peak NMS in the high laser power limit. The three peaks in the spectrum are a result of energy exchange between the mechanical and the two optical modes. We also demonstrate that the steady state bipartite entanglement between the three modes of the system can be efficiently controlled by the Kerr and optomechanical nonlinearity.  In addition, we have found that the photon tunneling can also control the optical, mechanical and entanglement properties of the system. Our results demonstrate that the present scheme can, in principle, be used as a sensitive optical switch/optical sensor. The mechanical and the two optical modes present in the two optically coupled microcavities coherently exchange energy and demonstrate tunable entanglement. This demonstrates that such a hybrid optomechanical system can be used to store and transfer information thus forming a part of a larger quantum information processing unit.

\begin{acknowledgements}
A. B. B acknowledges Birla Institute of Technology, Pilani for the facilities to carry out this research. S. M. acknowledges University of Petroleum and Energy Studies, Dehradun for providing the facilities to accomplish this research.
\end{acknowledgements}

\section{Appendix A}

All the input noise operators satisfies the following set of correlation functions \citep{100,101,102,103}

\begin{equation}\label{a1}
<a_{in}(t)a_{in}(t')>=<a_{in}^{\dagger}(t)a_{in}(t')>=0,
\end{equation}

\begin{equation}\label{a2}
<a_{in}(t)a_{in}^{\dagger}(t')>=\delta(t-t'),
\end{equation}

\begin{equation}\label{a3}
<b_{in}(t)b_{in}(t')>=<b_{in}^{\dagger}(t)b_{in}(t')>=0,
\end{equation}

\begin{equation}\label{a4}
<b_{in}(t)b_{in}^{\dagger}(t')>=\delta(t-t').
\end{equation}

The Brownian force noise operator of the mechanical mirror satisfies the following correlation function \citep{102}

\begin{equation}\label{a5}
<\zeta(t)\zeta(t')>=\frac{\gamma_{m}}{2\pi\omega_{m}}\int{\omega e^{-i\omega(t-t')}\left[1+\coth\left({\frac{\hbar\omega}{2k_{B}T}}\right)\right]d\omega},
\end{equation}

where, $k_{B}$ represents Boltzmann Constant and $T$ represents the temperature of the thermal bath connected to the mechanical mirror. Since the movable mirror attached to the thermal bath, therefore, there is random motion of the mirror that produces the Brownian noise. This kind of noise is Non-Markovian in nature \citep{100,101}.

For displacement spectrum, the correlation functions for various amplitude and phase noise quadratures and Brownian noise operator in Fourier space are given as \citep{103}:

\begin{equation}\label{a6}
<x_{1in}(\omega)x_{1in}(\Omega)>=2\pi\delta(\omega+\Omega)
\end{equation}

\begin{equation}\label{a7}
<y_{1in}(\omega)y_{1in}(\Omega)>=2\pi\delta(\omega+\Omega)
\end{equation}

\begin{equation}\label{a8}
<x_{1in}(\omega)y_{1in}(\Omega)>=2i\pi\delta(\omega+\Omega)
\end{equation}

\begin{equation}\label{a9}
<y_{1in}(\omega)x_{1in}(\Omega)>=-2i\pi\delta(\omega+\Omega)
\end{equation}

\begin{equation}\label{a10}
<x_{2in}(\omega)x_{2in}(\Omega)>=2\pi\delta(\omega+\Omega)
\end{equation}

\begin{equation}\label{a11}
<y_{2in}(\omega)y_{2in}(\Omega)>=2\pi\delta(\omega+\Omega)
\end{equation}

\begin{equation}\label{a12}
<x_{2in}(\omega)y_{2in}(\Omega)>=2i\pi\delta(\omega+\Omega)
\end{equation}

\begin{equation}\label{a13}
<y_{2in}(\omega)x_{2in}(\Omega)>=-2i\pi\delta(\omega+\Omega)
\end{equation}

\begin{equation}\label{a14}
<\zeta(\omega)\zeta(\Omega)>=2\pi\frac{\gamma_{m}}{\omega_{m}}\omega\left[1+\coth\left({\frac{\hbar\omega}{2k_{B}T}}\right)\right]\delta(\omega+\Omega).
\end{equation}

\section{Appendix B}

Using Routh-Hurwitz Criterion \citep{104}, the two stability conditions found for the system are as follows:

\begin{equation}\label{b1}
S_{1}=R_{0}>0,
\end{equation}

\begin{equation}\label{b2}
S_{2}=(R_{5}R_{4}R_{3}+R_{6}R_{1}R_{5}-R_{6}R_{3}^2-R_{2}R_{5}^2)>0,
\end{equation}

where,

\begin{eqnarray}\label{b3}
R_{0}&=&-k_{b}^2\Gamma_{1}^2\omega_{m}^2-k_{b}^2\omega_{m}^2\Delta_{1}^2+k_{b}^2\omega_{m}^2\delta_{1}^2+k_{b}^2k_{a}^2\omega_{m}^2-\omega_{m}k_{b}^2\delta_{1}a_{-}^2\nonumber\\
&+&\omega_{m}k_{b}^2\Delta_{1}a_{-}^2-2ik_{b}^2\Gamma_{1}\omega_{m}a_{-}a_{+}+2J^2\Gamma_{1}\omega_{m}^2k_{b}-\Gamma_{1}^2\omega_{m}^2\Delta_{b}^2\nonumber\\
&+&k_{a}^2\omega_{m}^2\Delta_{b}^2-2ia_{+}a_{-}\omega_{m}\Gamma_{1}\Delta_{b}^2+\Delta_{b}^2\Delta_{1}^2\omega_{m}^2-\Delta_{b}^2\delta_{1}^2\omega_{m}^2\nonumber\\
&-&\omega_{m}\Delta_{b}^2a_{+}(\Delta_{1}+\delta_{1})-a_{-}^2\Delta_{b}^2\delta_{1}\omega_{m}+a_{-}^2\Delta_{b}^2\Delta_{1}\omega_{m}+2J^2\Delta_{b}\delta_{1}\omega_{m}^2\nonumber\\
&+&J^2\Delta_{b}a_{-}^2\omega_{m}+J^2\Delta_{b}a_{+}^2\omega_{m}+2ia_{-}a_{+}\omega_{m}J^2k_{b}-J^4\omega_{m}^2,
\end{eqnarray}

\begin{eqnarray}\label{b4}
R_{1}&=&2k_{b}^2k_{a}\omega_{m}^2-2\Gamma_{1}^2\omega_{m}^2k_{b}-4ia_{+}a_{-}\omega_{m}\Gamma_{1}k_{b}+2k_{a}^2\omega_{m}^2k_{b}\nonumber\\
&-&\Gamma_{1}^2\gamma_{m}k_{b}^2+k_{b}^2k_{a}^2\gamma_{m}-\Delta_{1}^2\omega_{m}^2k_{b}+\delta_{1}^2\omega_{m}^2k_{b}+k_{b}^2\delta_{1}^2\gamma_{m}\nonumber\\
&-&k_{b}^2\Delta_{1}^2\gamma_{m}-k_{b}^2\Delta_{1}a_{+}^2-k_{b}^2\delta_{1}a_{+}^2-2\omega_{m}\delta_{1}a_{-}^2k_{b}+2\omega_{m}\Delta_{1}a_{-}^2k_{b}\nonumber\\
&+&2J^2\Gamma_{1}\gamma_{m}k_{b}-k_{b}\Gamma_{1}^2\omega_{m}^2-k_{b}\omega_{m}^2\Delta_{1}^2+k_{b}\omega_{m}^2\delta_{1}^2+2k_{a}J^2\omega_{m}^2\nonumber\\
&+&2k_{a}\omega_{m}^2\Delta_{b}^2+k_{a}^2\gamma_{m}\Delta_{b}^2-\Gamma_{1}^2\gamma_{m}\Delta_{b}^2\nonumber\\
&+&\Delta_{b}^2\Delta_{1}^2\gamma_{m}-\Delta_{b}^2\delta_{1}^2\gamma_{m}+2J^2\Delta_{b}\delta_{1}\gamma_{m}-J^4\gamma_{m},
\end{eqnarray}

\begin{eqnarray}\label{b5}
R_{2}&=&4k_{a}k_{b}\omega_{m}^2-k_{b}^2\Gamma_{1}^2+k_{b}^2k_{a}^2+2k_{b}^2k_{a}\gamma_{m}+k_{b}^2\omega_{m}^2\nonumber\\
&-&2\Gamma_{1}^2\gamma_{m}k_{b}+2k_{a}^2k_{b}\gamma_{m}-2\Delta_{1}^2\gamma_{m}k_{b}+2\delta_{1}^2\gamma_{m}k_{b}-2a_{+}^2\Delta_{1}k_{b}\nonumber\\
&-&2a_{+}^2\delta_{1}k_{b}-k_{b}^2\Delta_{1}^2+k_{b}^2\delta_{1}^2+2\Gamma_{1}J^2k_{b}-\Gamma_{1}^2\omega_{m}^2\nonumber\\
&-&2ia_{+}a_{-}\omega_{m}\Gamma_{1}+k_{a}^2\omega_{m}^2-\Delta_{1}^2\omega_{m}^2+\delta_{1}^2\omega_{m}^2-\omega_{m}\delta_{1}a_{-}^2\nonumber\\
&+&\omega_{m}\Delta_{1}a_{-}^2+2J^2k_{a}\gamma_{m}+2\omega_{m}^2J^2-\Delta_{b}^2\Gamma_{1}^2+k_{a}^2\Delta_{b}^2\nonumber\\
&+&2k_{a}\gamma_{m}\Delta_{b}^2+\omega_{m}^2\Delta_{b}^2+\Delta_{b}^2\Delta_{1}^2-\Delta_{b}^2\delta_{1}^2+2J^2\Delta_{b}\delta_{1}-J^4,
\end{eqnarray}

\begin{eqnarray}\label{b6}
R_{3}&=&2k_{b}^2k_{a}+2k_{a}^2k_{b}+4\gamma_{m}k_{b}k_{a}+\gamma_{m}k_{b}^2-2\Gamma_{1}^2k_{b}+2\omega_{m}^2k_{b}-2\Delta_{1}^2k_{b}\nonumber\\
&+&2\delta_{1}^2k_{b}+J^2k_{b}+2k_{a}\omega_{m}^2+k_{a}^2\gamma_{m}-\Delta_{1}^2\gamma_{m}+a_{+}^2\Delta_{1}\nonumber\\
&+&\delta_{1}^2\gamma_{m}-a_{+}^2\delta_{1}+2\gamma_{m}J^2+2k_{a}\Delta_{b}^2+\gamma_{m}\Delta_{b}^2,
\end{eqnarray}

\begin{eqnarray}\label{b7}
R_{4}&=&4k_{a}k_{b}+2\gamma_{m}k_{b}+k_{b}^2+2\gamma_{m}k_{a}-\Gamma_{1}^2\nonumber\\
&+&k_{a}^2+\omega_{m}^2-\Delta_{1}^2+\delta_{1}^2+J^2+\Delta_{b}^2,
\end{eqnarray}

\begin{equation}\label{b8}
R_{5}=2k_{b}+2k_{a}+\gamma_{m},
\end{equation}

\begin{equation}\label{b9}
R_{6}=1.
\end{equation}

\section{Appendix C}

The unknown coefficients used in Sec. IV are given as follows:

\begin{eqnarray}\label{c1}
X_{1}(\omega)&=&(\omega_m^2-i\omega\gamma_{m}-\omega^2)C_{1}(\omega)C_{6}(\omega)+ia_{+}\omega_{m}C_{3}(\omega)C_{6}(\omega)a_{-}\nonumber \\
&-&\omega_{m}C_{7}(\omega)C_{3}(\omega)[a_{+}(\Delta_{1}+\delta_{1})C_{3}(\omega)+ia_{-}C_{1}(\omega)-J^2\Delta_{b}a_{+}],
\end{eqnarray}

\begin{equation}\label{c2}
X_{2}(\omega)=J\sqrt{k_{b}}\omega_{m}[ia_{-}C_{6}(\omega)\Delta_{b}+C_{7}(\omega)[-(\Delta_{1}+\delta_{1})\Delta_{b}C_{3}(\omega)+C_{1}(\omega)(k_{b}-i\omega)+J^2\Delta_{b}^2]],
\end{equation}

\begin{equation}\label{c3}
X_{3}(\omega)=J\sqrt{k_{b}}\omega_{m}[iC_{6}(\omega)(k_{b}-i\omega)a_{-}+C_{7}(\omega)[-(\Delta_{1}+\delta_{1})(k_{b}-i\omega)C_{3}(\omega)-C_{1}(\omega)\Delta_{b}+J^2\Delta_{b}(k_{b}-i\omega)]],
\end{equation}

\begin{equation}\label{c4}
X_{4}(\omega)=\sqrt{k_{a}}\omega_{m}[-ia_{-}C_{3}(\omega)C_{6}(\omega)+C_{7}(\omega)C_{3}(\omega)[(\Delta_{1}+\delta_{1})C_{3}(\omega)-J^2\Delta_{b}]],
\end{equation}

\begin{equation}\label{c5}
X_{5}(\omega)=\sqrt{k_{a}}\omega_{m}C_{1}(\omega)C_{3}(\omega)C_{7}(\omega),
\end{equation}

\begin{equation}\label{c6}
X_{6}(\omega)=\omega_{m}\gamma_{m}\omega C_{6}(\omega)C_{6}(-\omega)\coth(\frac{\hbar\omega}{2k_{B}T}).
\end{equation}

Here,

\begin{equation}\label{c7}
C_{1}(\omega)=(\Gamma_{1}+k_{a}-i\omega)((k_{b}-i\omega)^2+\Delta_{b}^2)+J^2(k_{b}-i\omega),
\end{equation}

\begin{equation}\label{c8}
C_{2}(\omega)=(\delta_{1}-\Delta_{1})((k_{b}-i\omega)^2+\Delta_{b}^2)+J^2\Delta_{b},
\end{equation}

\begin{equation}\label{c9}
C_{3}(\omega)=(k_{b}-i\omega)^2+\Delta_{b}^2,
\end{equation}

\begin{equation}\label{c10}
C_{4}(\omega)=(\Gamma_{1}+k_{a}-i\omega)C_{1}(\omega)-(\Delta_{1}+\delta_{1})C_{2}(\omega),
\end{equation}

\begin{equation}\label{c11}
C_{5}(\omega)=(k_{b}-i\omega)C_{1}(\omega)+\Delta_{b}C_{2}(\omega),
\end{equation}

\begin{equation}\label{c12}
C_{6}(\omega)=C_{4}(\omega)C_{3}(\omega)+J^2C_{5}(\omega),
\end{equation}

\begin{equation}\label{c13}
C_{7}(\omega)=a_{+}C_{1}(\omega)-ia_{-}C_{2}(\omega).
\end{equation}

\end{document}